\documentclass[11pt]{article}
\usepackage{graphicx}
\usepackage{amsfonts}
\usepackage{theorem}
\newtheorem{Lem}{Lemma}[section]
\newtheorem{Def}[Lem]{Definition}
\newtheorem{The}[Lem]{Theorem}

\newtheorem{Cor}[Lem]{Corollary}

\newtheorem{axiom}[Lem]{Axiom}

\newtheorem{Rem}[Lem]{Remark}

\newcommand{\qed}{\hbox{\rule{6pt}{6pt}}}

\begin{document}
\title{A generalized Fannes' inequality}
\author{S. Furuichi$^1$\footnote{E-mail:furuichi@chs.nihon-uac.jp}, K.Yanagi$^2$\footnote{E-mail:yanagi@yamaguchi-u.ac.jp} and K.Kuriyama$^2$\footnote{E-mail:kuriyama@yamaguchi-u.ac.jp}\\
$^1${\small Department of Computer Science and System Analysis, College
of Humanities and Sciences, 
Nihon University,}\\{\small 3-25-40, Sakurajyousui, Setagaya-ku, Tokyo, 156-8550, Japan}\\
$^2${\small Division of Applied Mathematical Science, Graduate School of Science and Engineering,} \\{\small 
Yamaguchi University, Tokiwadai 2-16-1, Ube City, 755-0811, Japan}}
\date{}
\maketitle

{\bf Abstract.} 
We axiomatically characterize the Tsallis entropy of a finite quantum system.
In addition, we derive a continuity property of Tsallis entropy. This gives a generalization of the Fannes' inequality.
\vspace{3mm}


\vspace{3mm}


\vspace{3mm}

\section{Introduction with uniqueness theorem of Tsallis entropy}
Three or four decades ago, some extensions of Shannon entropy was studied by many researchers \cite{AD}. 
In statistical physics, the Tsallis entropy was defined in \cite{Tsa} by 
$$H_q(X) \equiv \frac{\sum_{x}\left( p(x)^q -p(x) \right)}{1-q} = \sum_x\eta_q\left(p(x)\right)$$ 
with one parameter $q \in \mathbf{R}^+$ as an extension of Shannon entropy $H_1(X) = -\sum_x p(x)\log p(x)$, for any probability distribution  
$p(x) \equiv p(X=x)$ of a given random variable $X$. Where $q$-entropy function is defined by $\eta_q(x)\equiv -x^q \ln_qx =\frac{x^q-x}{1-q}$ 
and the $q$-logarithmic function $\ln_qx \equiv \frac{x^{1-q} -1}{1-q}$ is defined for $q \geq 0$, $q \neq 1$ and $x \geq 0$.  

The Tsallis entropy $H_q(X)$ converges to the Shannon entropy $-\sum_{x} p(x) \log p(x)$ as $q \to 1$. 
See \cite{FYK} for fundamental properties of the Tsallis entropy and the Tsallis relative entropy.
In the previous paper \cite{Furu}, we gave the uniqueness theorem for the Tsallis entropy for a classical system, introducing the generalized Faddeev's axiom.
We briefly review the uniqueness theorem for Tsallis entropy below.

The function $I_q(x_1,\cdots ,x_n)$ is assumed to be defined on $n$-tuple $(x_1, \cdots ,x_n)$ belonging to 
$$\Delta _n \equiv \left\{ (p_1,\cdots ,p_n) \vert \sum_{i=1}^n p_i =1, p_i \geq 0\,\, (i=1,2,\cdots ,n)\right\}$$ and to take values in $\mathbf{R}^+ \equiv [0,\infty )$.
Then we adopted the following generalized Faddeev's axiom.
\begin{axiom}{\bf (Generalized Faddeev's axiom)}\label{gfaddeev}
\begin{itemize}
\item[(F1)] {\it Continuity}: The function $f_q (x) \equiv I_q(x,1-x)$ with parameter $q\geq 0$ is continuous on the closed interval $\left[0,1\right]$
 and $f_q(x_0) > 0$ for some  $x_0 \in \left[0,1\right]$. 
\item[(F2)] {\it Symmetry}: 
For arbitrary permutation $\left\{x'_k\right\} \in \Delta_n$ of $\left\{x_k\right\} \in \Delta_n$,
\begin{equation}
I_q(x_1,\cdots ,x_n) = I_q(x'_1,\cdots ,x'_n).
\end{equation}
\item[(F3)] {\it Generalized additivity}:
For $x_n= y +z$, $y \geq 0$ and $z >0$,
\begin{equation}
I_q(x_1,\cdots ,x_{n-1},y,z) = I_q(x_1,\cdots ,x_n)  + x_n^q I_q \left(\frac{y}{x_n},\frac{z}{x_n}\right).
\end{equation}
\end{itemize}
\end{axiom}

\begin{The} {\bf (\cite{Furu})}  \label{the} 
The conditions (F1), (F2) and (F3) uniquely give the form of the function $I_q : \Delta_n \to \mathbf{R}^+$ such that
\begin{equation} \label{the_eq0}
I_q(x_1,\cdots ,x_n) = \mu_q H_q(x_1,\cdots ,x_n),
\end{equation}
where $\mu_q$ is a positive constant that depends on the parameter $q>0$.
\end{The}

If we further impose the normalized condition on Theorem \ref{the}, it determines the entropy of type $\beta$
(the structural $a$-entropy), (see p.189 on \cite{AD}).

\begin{Def}
For a density operator $\rho$ on a finite dimensional Hilbert space $\mathbf{H}$, the Tsallis entropy is defined by
$$
S_q(\rho) \equiv \frac{\hbox{Tr}[\rho^q -\rho]}{1-q} =\hbox{Tr}[\eta_q(\rho)],
$$ 
with a nonnegative real number $q$. 
\end{Def}

Note that Tsallis entropy is a particular case of $f$-entropy \cite{Weh}. See also \cite{Petz} for a quasi-entropy which 
is a quantum version of $f$-divergence \cite{Csi}.

Let $T_q$ be a mapping on the set $S(\mathbf{H})$ of all density operators to $\mathbf{R}^+$. 
\begin{axiom}  \label{axiom1}
We give the postulates which the Tsallis entropy should satisfy. 
\begin{itemize}
\item[(T1)] {\it Continuity:} For $\rho \in S(\mathbf{H})$, $T_q(\rho)$ is a continuous function with respect to the $1$-norm $\left\|\cdot \right\|_1$.
\item[(T2)] {\it Invariance:} For unitary transformation $U$, $T_q(U^*\rho U) = T_q(\rho)$.
\item[(T3)] {\it Generalized mixing condition:} For $\rho = \mathop  \oplus \limits_{k = 1}^n \lambda_k\rho _k$ on $\mathbf{H} = \mathop  \oplus \limits_{k = 1}^n \mathbf{H} _k$, where 
$\lambda_k \geq 0, \sum_{k=1}^n \lambda_k =1, \rho_k \in S(\mathbf{H}_k)$, we have the additivity:
$$T_q(\rho) = \sum_{k=1}^n \lambda_k^qT_q(\rho_k) + T_q( \lambda_1,\cdots,\lambda_n),$$
where $(\lambda_1,\cdots,\lambda_n) $ represents the diagonal matrix $(\lambda_k \delta_{kj})_{k,j=1,\cdots,n}$.
\end{itemize}
\end{axiom}

\begin{The}
If $T_q$ satisfies Axiom \ref{axiom1}, then $T_q$ is uniquely given by the following form
$$ T_q(\rho) = \mu_q S_q(\rho), $$ 
with a positive constant number $\mu_q$ depending on parameter $q > 0$.
\end{The}
{\it Proof}:
Although the proof is quite similar to that of Theorem 2.1 in \cite{OP}, we give it for readers' convenience. 
From (T2) and (T3), we have
$$T_q(\lambda_1,\lambda_2)=\lambda_1^q T_q(1) + \lambda_2^qT_q(1) + T_q(\lambda_1,\lambda_2), $$
which implies $T_q(1) = 0$.
Moreover, by (T2) and (T3), when $p_n \neq 1$, we have

\begin{eqnarray*}
 T_q \left( {p_1 , \cdots ,p_{n - 1} ,\lambda p_n ,\left( {1 - \lambda } \right)p_n } \right) &=& p_n ^q T_q \left( {\lambda ,1 - \lambda } \right) 
  + \left( {1 - p_n } \right)^q T_q \left( {\frac{{p_1 }}{{1 - p_n }}, \cdots ,\frac{{p_{n - 1} }}{{1 - p_n }}} \right)\\
 &+& T_q \left( {p_n ,1 - p_n } \right) 
 \end{eqnarray*}
and 
\[
T_q \left( {p_1 , \cdots ,p_{n - 1} ,p_n } \right) = 
p_n ^q T_q \left( 1 \right) + \left( {1 - p_n } \right)^q T_q \left( {\frac{{p_1 }}{{1 - p_n }}, \cdots ,\frac{{p_{n - 1} }}{{1 - p_n }}} \right) 
+ T_q \left( {p_n ,1 - p_n } \right).
\]
From these equations, we have
\begin{equation}   \label{eq_theorem_01}
T_q \left( {p_1 , \cdots ,p_{n - 1} ,\lambda p_n ,\left( {1 - \lambda } \right)p_n } \right) 
= T_q \left( {p_1 , \cdots ,p_{n - 1} ,p_n } \right) + p_n^q T_q \left( {\lambda ,1 - \lambda } \right)
\end{equation}
If we set $\lambda p_n = y$ and $(1-\lambda) p_n =z$ in Eq.(\ref{eq_theorem_01}), then for $p_n=y+z \neq 0$ we have 
\begin{equation}
T_q \left( {p_1 , \cdots ,p_{n - 1} ,y ,z } \right) 
= T_q \left( {p_1 , \cdots ,p_{n - 1} ,p_n } \right) + p_n^q T_q \left( {\frac{y}{p_n} ,\frac{z}{p_n} } \right).
\end{equation}
Then for any $x,y,z \in \mathbf{R}$ such that $0\leq x,y <1$, $0< z \leq 1$ and $x+y+z=1$, we have
\begin{eqnarray*}
T_q(x,y,z) &=& T_q(x,y+z) +(y+z)^q T_q\left(\frac{y}{y+z},\frac{z}{y+z}\right)\\
&=&T_q(y,x+z)+(x+z)^qT_q\left(\frac{x}{x+z},\frac{z}{x+z}\right).
\end{eqnarray*}
If we set $t_q(x)\equiv T_q(x,1-x)$, then we have
$$t_q(x)+(1-x)^qt_q\left(\frac{y}{1-x}\right)=t_q(y)+(1-y)^qt_q\left(\frac{x}{1-y}\right).$$
Taking $x=0$ and some $y>0$, we have $T_q(0,1) = t_q(0) = 0$ for $q \neq 0$.
Again setting $\lambda = 0$ in Eq.(\ref{eq_theorem_01}) and using (T2), we have the reducing condition
$$T_q(p_1,\cdots,p_n,0) = T_q(p_1,\cdots,p_n).$$
Thus $T_q$ satisfies all conditions of the generalized Faddeev's axiom (F1), (F2) and (F3).
Therefore we can apply Theorem \ref{the} so that we obtain $T_q(\lambda_1,\cdots,\lambda_n) = \mu_q H_q(\lambda_1,\cdots,\lambda_n)$.
Thus we have $ T_q(\rho) = \mu_q S_q(\rho), $ for density operator $\rho$.
\hfill \qed

\begin{Rem}
For the special case $q=0$ in the above theorem, we need the reducing condition as an additional axiom.
\end{Rem}

\section{A continuity of Tsallis entropy}

We give a continuity property of the Tsallis entropy $S_q(\rho)$. To do so, we state a few lemmas.

\begin{Lem} \label{lemma1}
For a density operator $\rho$ on the finite dimensional Hilbert space $\mathbf{H}$, we have
$$S_q(\rho) \leq \ln_q d,$$
where $d = \dim \mathbf{H} < \infty $.
\end{Lem}
{\it Proof}:
Since we have $\ln_qz \leq z-1$ for $q \geq 0$ and $z \geq 0$, we have $\frac{x-x^{q}y^{1-q}}{1-q} \geq x-y$ for $x \geq 0$, $y \geq 0$, $q \geq 0$ and $q \neq 1$,
Therefore the Tsallis relative entropy \cite{FYK}:
$$
D_q(\rho\vert \sigma) \equiv \frac{\hbox{Tr}[\rho -\rho^q\sigma^{1-q}]}{1-q}
$$
for two commuting density operators $\rho$ and $\sigma$, $q \geq 0$ and $q \neq 1$,
 is nonnegative. Then we have
$0 \leq D_q(\rho\vert \frac{1}{d}I) = -d^{q-1}\left(S_q(\rho) - \ln_qd\right)$.
Thus we have the present lemma.
\hfill \qed

\begin{Lem} \label{lemma2}
If $f$ is a concave function and $f(0) = f(1) =0$, then we have 
$$ \vert f(t+s) - f(t) \vert \leq \max \left\{f(s), f(1-s)\right\}$$
for any $s \in [0,1/2]$ and $t \in [0,1]$ satisfying $0 \leq s + t \leq 1$.
\end{Lem}
{\it Proof}:
\begin{itemize}
\item[(1)] Consider the function $r(t) = f(s)-f(t+s)+f(t)$. Then $r'(t) \geq 0$ since $f'$ is a monotone decreasing function. 
Thus we have $r(t) \geq 0$ by $r(0)=0$. Therefore
$f(t+s) -f(t) \leq f(s).$
\item[(2)] Consider the function of $l(t) = f(t+s) -f(t) +f(1-s)$. Then $l'(t) \leq 0$. Thus we have $l(t) \geq 0$ by $l(1-s)=0$.
Therefore $-f(1-s) \leq f(t+s) - f(t)$.
\end{itemize}
Thus we have the present lemma.
\hfill \qed

\begin{Lem} \label{lemma3}
For any real number $u, v \in [0,1]$ and $q \in [0,2]$,
if $\vert u -v \vert \leq \frac{1}{2} $, then $\vert \eta_q(u) - \eta_q(v) \vert \leq \eta_q (\vert u - v \vert)$.
\end{Lem}
{\it Proof}:
Since $\eta_q$ is a concave function with $\eta_q(0)=\eta_q(1)=0$, we have
$$\vert \eta_q(t+s) - \eta_q(t) \vert \leq \max\left\{\eta_q(s),\eta_q(1-s) \right\}$$
for $s \in [0,1/2]$ and $t \in [0,1]$ satisfying $ 0 \leq t+s \leq 1$, by Lemma \ref{lemma2}.
Here we set 
$$h_q(s)\equiv \eta_q(s)-\eta_q(1-s),\quad s \in [0,1/2],\,\,\, q \in [0,2].$$
Then we have $h_q(0)=h_q(1/2)=0$ and $h''_q(s) \leq 0$ for $s \in [0,1/2]$. Therefore we have $h_q(s) \geq 0$, 
which implies
$$\max\left\{\eta_q(s),\eta_q(1-s) \right\} =    \eta_q(s).$$
Thus we have the present lemma by letting $u = t+ s$ and $v = t$.
\hfill \qed

\begin{The} \label{1theorem}
For two density operators $\rho_1$ and $\rho_2$ on the finite dimensional Hilbert space $\mathbf{H}$ with $\dim \mathbf{H} =d$ and $q \in [0,2]$, 
if $\left\| \rho_1 -\rho_2 \right\|_1 \leq q^{1/(1-q)}$, then 
$$\vert S_q(\rho_1) - S_q(\rho_2) \vert \leq \left\| \rho_1 -\rho_2 \right\|_1^q \ln_q d+ \eta_q(\left\| \rho_1 -\rho_2 \right\|_1).$$ 
Where we denote $\left\| A \right\|_1 \equiv \hbox{Tr} \left[  (A^*A)^{1/2} \right]$ for a bounded linear operator $A$.
\end{The}
{\it proof}:
Let $\lambda_1^{(1)} \geq \lambda_2^{(1)} \geq \cdots \geq \lambda_d^{(1)}$ and $\lambda_1^{(2)} \geq \lambda_2^{(2)} \geq \cdots \geq \lambda_d^{(2)}$
be eigenvalues of two density operators $\rho_1$ and $\rho_2$, respectively.  (The degenerate eigenvalues are repeated according to their multiplicity.)
We set $\varepsilon \equiv \sum_{j=1}^d \varepsilon_j$ and $\varepsilon_j \equiv \vert \lambda_j^{(1)} - \lambda_j^{(2)} \vert$.
Then we have
$$\varepsilon_j \leq \varepsilon \leq \left\| \rho_1 - \rho_2\right\|_1 \leq q^{1/(1-q)} \leq 1/2$$
by Lemma 1.7 of \cite{OP}.
Applying Lemma \ref{lemma3}, we have
$$\vert S_q(\rho_1) - S_q(\rho_2) \vert \leq \sum_{j=1}^d \vert \eta_q(\lambda_j^{(1)}) - \eta_q(\lambda_j^{(2)}) \vert \leq \sum_{j=1}^d \eta_q(\varepsilon_j).$$ 
By the formula $\ln_q(xy) = \ln_qx + x^{1-q} \ln_q y$, we have
\begin{eqnarray*}
\sum_{j=1}^d \eta_q(\varepsilon_j) &=& - \sum_{j=1}^d \varepsilon_j^q \ln_q \varepsilon_j = \varepsilon \left\{-\sum_{j=1}^d \frac{\varepsilon_j^q}{\varepsilon} \ln_q\left(\frac{\varepsilon_j}{\varepsilon}\varepsilon\right) \right\} \\
&=& \varepsilon \left\{   - \sum_{j=1}^d \frac{\varepsilon_j^q}{\varepsilon}\ln_q \frac{\varepsilon_j}{\varepsilon}  - \sum_{j=1}^d \frac{\varepsilon_j^q}{\varepsilon}  
 \left(\frac{\varepsilon_j}{\varepsilon}  \right)^{1-q}  \ln_q \varepsilon   \right\} \\
&=& \varepsilon^q \sum_{j=1}^d \eta_q\left(\frac{\varepsilon_j}{\varepsilon}  \right) + \eta_q(\varepsilon) \\
& \leq & \varepsilon^q \ln_q d + \eta_q(\varepsilon).
\end{eqnarray*}
In the above inequality, Lemma \ref{lemma1} was used for $\rho=\left(\varepsilon_1/\varepsilon,\cdots,\varepsilon_d/\varepsilon\right)$. Therefore we have
$$\vert S_q(\rho_1) - S_q(\rho_2) \vert \leq \varepsilon^q \ln_q d + \eta_q(\varepsilon). $$
Now $\eta_q(x)$ is a monotone increase function on $x \in [0,q^{1/(1-q)}]$.
In addition, $x^q$ is a monotone increasing function for $q \in [0,2]$. Thus we have the present theorem.
\hfill \qed

By taking the limit as $q \to 1$, we have the following Fannes' inequality (see pp.512 of \cite{NC}, also \cite{Fan,AF,OP}) as a cororally, since $\lim_{q\to 1} q^{1/(1-q)} = \frac{1}{e}$.
\begin{Cor}
For two density operators $\rho_1$ and $\rho_2$ on the finite dimensional Hilbert space $\mathbf{H}$ with $\dim \mathbf{H} = d < \infty$,
 if $\left\| \rho_1 -\rho_2 \right\|_1 \leq \frac{1}{e} $, then  
$$\vert S_1(\rho_1) - S_1(\rho_2) \vert \leq \left\| \rho_1 -\rho_2 \right\|_1 \ln d+ \eta_1(\left\| \rho_1 -\rho_2 \right\|_1),$$ 
where $S_1$ represents the von Neumann entropy $S_1(\rho) = \hbox{Tr}[\eta_1(\rho)]$ and $\eta_1(x) = -x \ln x$.
\end{Cor}
\section*{Acknowledgement}
The authors would like to thank referees for careful reading and providing valuable comments to improve the manuscript.
The author (S.F.) was partially supported by the Japanese Ministry of Education, Science, Sports and Culture, Grant-in-Aid for 
Encouragement of Young scientists (B), 17740068.

\end{document}